\documentclass[12pt]{iopjournal}
\usepackage[utf8]{inputenc}
\usepackage[T1]{fontenc}
\usepackage{amsmath,amssymb}
\usepackage{lineno}
\usepackage{graphicx}
\usepackage{url}
\usepackage{ragged2e}
\usepackage{xcolor}

\begin{document}
\articletype{Paper}

\title{Transformer-Based Pulse Shape Discrimination in HPGe Detectors with Masked Autoencoder Pre-training}

\author{Marta Babicz$^{1,*}$, Saúl Alonso-Monsalve$^{2}$, Alain Fauquex$^{1,2}$ and Laura Baudis$^{1}$}

\affil{$^1$University of Zürich, Physik-Institut, Winterthurerstrasse 190, Zürich 8057, Switzerland.}

\affil{$^2$ETH Zürich, Rämistrasse 101, Zürich 8092, Switzerland.}

\affil{$^*$Author to whom any correspondence should be addressed.}

\email{marta.babicz@physik.uzh.ch}

\begin{abstract}
\justifying
Pulse-shape discrimination (PSD) in high-purity germanium (HPGe) detectors is central to rare-event searches such as neutrinoless double-beta decay ($0\nu\beta\beta$), yet conventional approaches compress each waveform into a small set of summary parameters, potentially discarding information in the full time series that is relevant for classification. We benchmark transformer-based models that operate directly on digitised waveforms using the \textsc{Majorana Demonstrator} AI/ML data release. Models are trained to reproduce the collaboration-provided accept/reject labels for four standard PSD cuts and to regress calibrated energy. We compare supervised training from scratch, masked autoencoder (MAE) self-supervised pre-training followed by fine-tuning, and a feature-based gradient-boosted decision tree (GBDT) baseline. Transformers outperform GBDT across all PSD targets, with the largest gains on the most challenging labels and on the combined PSD-pass definition. MAE pre-training improves sample efficiency, reducing labelled-data requirements by factors of 2–4 in low-label regimes. For energy regression, both transformer variants show a small common underestimation on the test split, while fine-tuning modestly narrows the residual distribution. These results motivate follow-up studies of robustness across detectors and operating conditions and of performance near $Q_{\beta\beta}$.
\end{abstract}

\section{Introduction}
\justifying

High-purity germanium (HPGe) detectors enriched in ${}^{76}$Ge are a leading technology for neutrinoless double-beta decay ($0\nu\beta\beta$) searches because the isotope serves simultaneously as the decay source and the active detection medium, enabling high detection efficiency together with excellent energy resolution and operation in ultra-low-background environments. Because any $0\nu\beta\beta$ signal is expected to be exceedingly rare, with half-lives $\gtrsim 10^{26}$~yr, maximising sensitivity requires aggressive background suppression and highly efficient event classification. A discovery would establish lepton-number violation ($\Delta L=2$) and imply Majorana neutrinos~\cite{Dolinski2019Review}.

To date, experiments such as \textsc{GERDA}~\cite{Agostini2023GERDA} and the \textsc{Majorana Demonstrator} (MJD)~\cite{majorana_construction_commissioning_performance,MJD_final_results} have set stringent limits on the $0\nu\beta\beta$ half-life in ${}^{76}$Ge, translating, under standard nuclear-physics assumptions, into constraints on the effective Majorana mass $m_{\beta\beta}$.
The staged \textsc{LEGEND} program builds on the achievements of \textsc{GERDA} and the \textsc{Majorana Demonstrator}: \textsc{LEGEND}-200 has already reported first $0\nu\beta\beta$ search results \cite{LEGEND200_first_results} and continues data taking, while \textsc{LEGEND}-1000 is the planned ton-scale phase currently under preparation~\cite{big_legend_paper}. This motivates event-classification techniques that exploit the full information content of digitised detector waveforms for both ongoing \textsc{LEGEND}-200 analyses and the longer-term \textsc{LEGEND}-1000 program.

Pulse-shape discrimination (PSD) is a primary handle for background rejection in HPGe $0\nu\beta\beta$ searches. Conventional analyses use a small set of physically motivated waveform summary parameters to define acceptance regions that efficiently suppress dominant background populations~\cite{Agostini2023GERDA,MJD_final_results}. This approach is robust, but it can discard information contained in the full waveform time series, motivating end-to-end machine-learning models that learn representations directly from digitised signals.

Supervised learning in this setting is complicated by the lack of event-by-event topology labels. Instead, training targets typically rely on simulations and detector-response models, or on analysis-defined calibration proxies, both of which may be imperfect and can introduce label noise or domain mismatch. Self-supervised pre-training methods such as masked autoencoding can exploit abundant unlabelled waveforms to learn transferable representations, reducing the amount of labelled or proxy-labelled data required for downstream PSD tasks.
We quantify these sample-efficiency gains by subsampling the labelled data to emulate low-label regimes relevant both for ongoing \textsc{LEGEND}-200 analyses and for the future \textsc{LEGEND}-1000 program.

In this work, we evaluate transformer-based PSD using waveforms from MJD and address three questions: (1) whether direct waveform modelling with detector-conditioned transformers improves agreement with analysis-defined PSD labels relative to feature-based methods; (2) whether masked autoencoder (MAE) pre-training on unlabelled calibration waveforms improves sample efficiency for downstream PSD tasks; and (3) how these benefits depend on supervised training-set size and training duration. Our main contributions are:

\begin{itemize}
  \item We develop a detector-conditioned transformer architecture for HPGe waveforms that operates directly on digitised charge traces and their gradients, avoiding hand-crafted feature compression while allowing adaptation across detectors.
  \item We show that masked autoencoder self-supervised pre-training on unlabelled waveforms improves sample efficiency for downstream PSD tasks, reducing the amount of labelled data needed to reach a given performance level, especially in low-label regimes.
  \item We benchmark against a feature-based gradient-boosted decision tree baseline and find that the largest gains from end-to-end waveform modelling and pre-training appear for the most challenging PSD targets and for the combined PSD-pass definition.
  \item We additionally study energy regression as an auxiliary task and find that fine-tuning from the pre-trained encoder modestly narrows the residual distribution relative to supervised training from scratch.
\end{itemize}

\subsection{High-Purity Germanium Detectors}\label{sec:HPGeDetectors}
In the \textsc{Majorana Demonstrator}, cm-scale HPGe detector diodes are fabricated from germanium enriched to $\simeq 88$--$90\%$ in ${}^{76}$Ge~\cite{majorana_construction_commissioning_performance}, providing the $0\nu\beta\beta$ source isotope directly within the active volume. The diodes are operated under reverse bias and fully depleted. When ionising radiation deposits energy in the depleted bulk, electron--hole pairs are created in approximate proportion to the deposited energy, with a mean pair-creation energy of $\sim 2.96~\mathrm{eV}$ in Ge at cryogenic temperatures. The electrons and holes drift in opposite directions towards the $n^+$ and $p^+$ contacts under the influence of the electric field.

The measured signal is governed not by charge \emph{arrival} at an electrode, but by the current \emph{induced} on the readout electrode as charge carriers move through the device, as described by the Shockley--Ramo theorem~\cite{ShockleyPaper,RamoPaper}. The instantaneous induced current on electrode $A$ from a charge $q$ at position $\vec{x}(t)$ with velocity $\vec{v}(t)$ is
\begin{align}
    i_A(t) = q\,\vec{v}(t)\cdot \vec{E}_{w,A}(\vec{x}(t))
           = -q\,\frac{\mathrm{d}\phi_{w,A}(\vec{x}(t))}{\mathrm{d}t},
    \label{eq:ramo1}
\end{align}
where $\phi_{w,A}$ is the \emph{weighting potential} of electrode $A$ and $\vec{E}_{w,A}=-\nabla\phi_{w,A}$ is the corresponding \emph{weighting field}. The weighting potential $\phi_{w,A}$ is obtained by solving Laplace's equation in the detector volume with $\phi_{w,A}=1$ on electrode $A$ and $\phi_{w,A}=0$ on all other electrodes (with space charge set to zero for this auxiliary problem).

Integrating Eq.~\eqref{eq:ramo1} gives the induced charge on electrode $A$. For a single electron--hole pair, the total induced charge is the sum of the electron and hole contributions,
\begin{align}
Q_A(t)=(-q)\left[\phi_{w,A}(\vec{x}_e(t))-\phi_{w,A}(\vec{x}_e(0))\right]
      +(+q)\left[\phi_{w,A}(\vec{x}_h(t))-\phi_{w,A}(\vec{x}_h(0))\right],
\label{eq:ramoQ_pair}
\end{align}
where $\vec{x}_e(t)$ and $\vec{x}_h(t)$ are the electron and hole positions, carrying charges $-q$ and $+q$, respectively. For a pair created at a common initial position $\vec{x}_0$ (so that $\vec{x}_e(0)=\vec{x}_h(0)=\vec{x}_0$), Eq.~\eqref{eq:ramoQ_pair} can be written as
\begin{align}
Q_A(t)=q\left[\phi_{w,A}(\vec{x}_h(t))-\phi_{w,A}(\vec{x}_e(t))\right],
\end{align}
up to an additive constant that depends on the choice of reference for induced charge at $t=0$; only differences in induced charge are physically relevant.

\begin{figure}[ht!]
    \centering
    \includegraphics[scale=0.6]{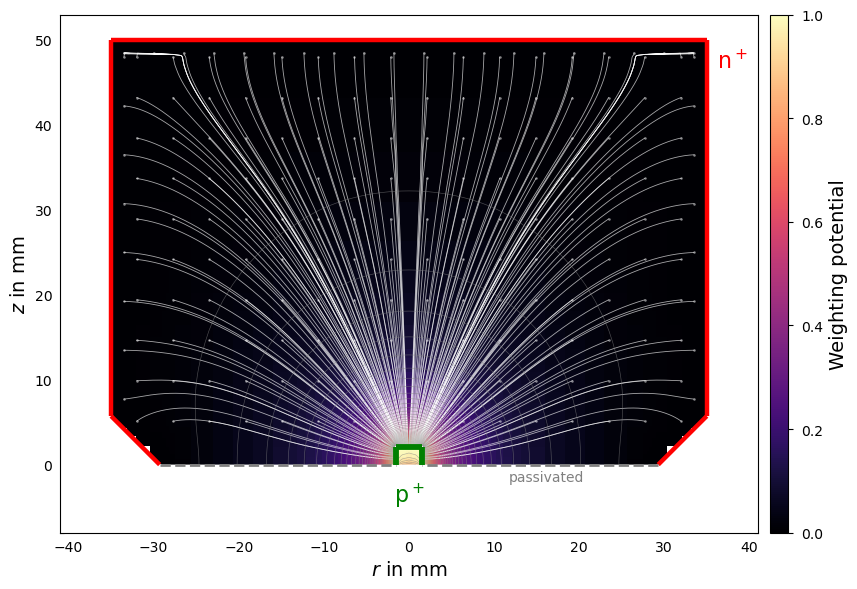}
    \caption{Cross-section of a Majorana Demonstrator--style p-type point-contact (PPC) HPGe detector simulated with \texttt{SolidStateDetectors.jl} (SSD)~\cite{Abt:2021SSD}. The colormap shows the point-contact weighting potential $\phi_w(r,z)$, normalised such that $\phi_w=1$ at the p$^{+}$ point contact (green) and $\phi_w=0$ at the outer n$^{+}$ Li-diffused contact (red); grey contours indicate $\phi_w$ equipotentials. White curves show electric-field lines obtained from the SSD field solution under the applied bias (interpretable as hole drift trajectories, which follow $\vec{E}$ toward the p$^{+}$ contact; electrons drift oppositely toward the n$^{+}$ mantle). The dashed grey segment marks the passivated bottom surface between the contacts. Axes show the cylindrical-symmetry coordinates $r$ and $z$ in mm.}
    \label{fig:weighting}
\end{figure}

The HPGe detectors used in MJD are p-type point-contact (PPC) detectors (Fig.~\ref{fig:weighting}). In PPC geometries, the weighting potential of the readout $p^{+}$ contact is sharply localised near the point contact. As a result, bulk events typically exhibit a current peak when carriers enter this high-weighting-field region; the signal is usually dominated by hole motion near the point contact, while electrons drift to the $n^{+}$ mantle where $\phi_w \approx 0$.

A lithium-diffused $n^+$ contact surrounds most of the outer detector surface and forms a dead layer (typically of order $0.5$--$1\,\mathrm{mm}$) with reduced or negligible charge collection, which suppresses sensitivity to surface $\alpha$ particles and many external $\beta$ backgrounds on that surface~\cite{Aguayo2013NIMAPPC,Arnquist2022AlphaRejection}. Between the fully active bulk and the dead layer, detectors can exhibit a \emph{transition region} with partial and/or delayed charge collection. In contrast, the surface region near the $p^+$ point contact (including the passivated surface) can show anomalous or delayed charge collection, leading to distinctive pulse-shape features and motivating dedicated surface-event rejection cuts in analysis.

\subsection{Traditional pulse-shape discrimination in the \textsc{Majorana Demonstrator}}
\label{sec:TraditionalPSD}
In MJD, PSD is implemented using engineered scalar quantities derived from the digitised charge waveform (and corresponding current estimators) to suppress key background classes, including multi-site $\gamma$ interactions, surface $\alpha$ events, and events with partial or delayed charge collection. The PPC geometry, with a weighting potential sharply peaked near the point contact, enhances sensitivity to interaction topology and near-surface charge-collection effects, enabling efficient selection cuts.

In this work, we use four binary accept/reject PSD labels provided in the MJD AI/ML data release~\cite{MD_Data_Release} as reference targets: low-side AvsE, high-side AvsE, DCR, and LQ. These labels are analysis-defined proxies for underlying event classes rather than direct event-by-event \textit{physics-truth} topology labels.

\subsubsection{AvsE selections}

\begin{figure}[htb]
  \centering
  \hspace*{-1.0cm}
  \includegraphics[width=1.0\linewidth]{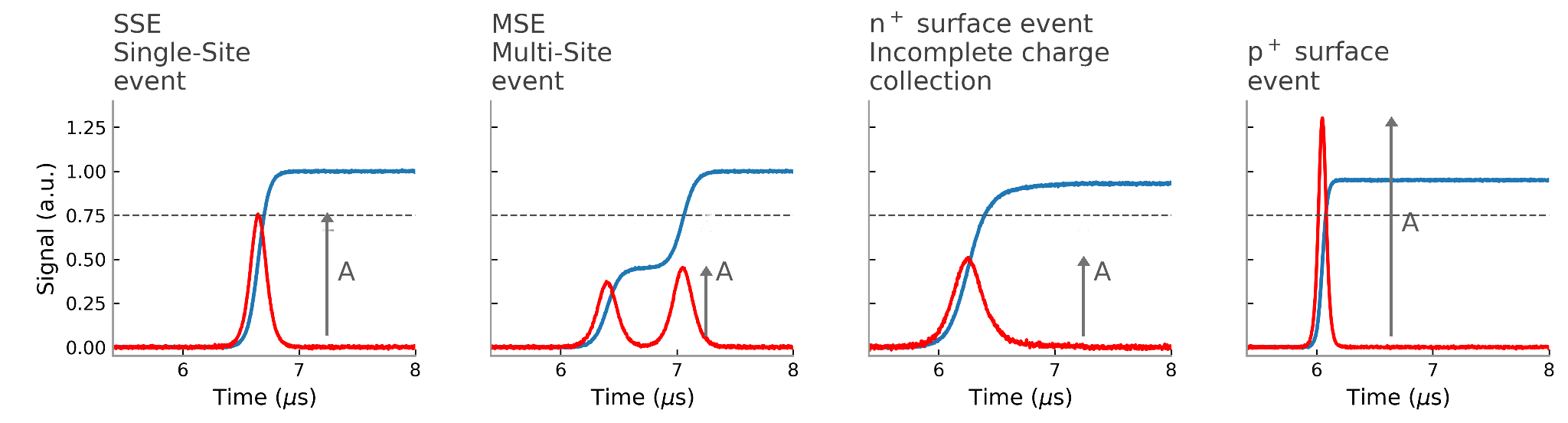}
  \par\vspace{-.25cm}
  \caption{Illustrative charge-waveform (blue) and current-estimator (red) traces motivating the AvsE selections. AvsE uses the maximum current-estimator amplitude $A$ relative to energy $E$ to separate SSE-like events from MSE-like $\gamma$ events, which can exhibit multiple current peaks and typically smaller corrected AvsE.}
  \label{fig:cuts}
\end{figure}

AvsE is a discriminator constructed from a maximum induced-current (or current-estimator) amplitude $A$ and the calibrated energy $E$, with detector- and drift-time-dependent corrections applied in the MJD analysis~\cite{MJD_final_results}. Events that are single-site-like (SSE-like), as expected for $0\nu\beta\beta$ at detector resolution, typically exhibit a single dominant current peak and populate a characteristic band in AvsE. Multi-site $\gamma$ events (MSEs) can produce multiple spatially separated energy depositions, leading to multi-lobed current signatures and, on average, smaller corrected AvsE values; these are rejected by a low-side AvsE selection. Events very near the point contact can yield unusually large AvsE due to the rapid final drift through the high-weighting-field region and are rejected by a high-side AvsE selection. Figure~\ref{fig:cuts} shows representative charge-waveform and current-estimator traces that motivate the low-side and high-side AvsE selections.

\subsubsection{LQ selection}
The late-charge (LQ) parameter~\cite{MJD_final_results} targets events with partial and/or delayed charge collection associated with the transition region between the outer $n^{+}$ contact and the fully active bulk. Such events can exhibit a fast initial rise followed by additional late charge, distorting the pulse shape and biasing the reconstructed energy. In the MJD analysis, LQ is defined using the charge waveform segment after it reaches 80\% of its final amplitude, quantifying excess late charge relative to an idealised prompt-collection trajectory; a high-side selection rejects events with anomalously large late-charge contributions.

\subsubsection{DCR selection}
The delayed charge recovery (DCR) parameter rejects surface $\alpha$ interactions that produce delayed charge-collection components, which can appear as an anomalous slope in the late-time tail of the processed charge waveform~\cite{MJD_final_results,DCR_Cut_Paper}. We do not recompute DCR here; we use the provided accept/reject label derived from the corrected DCR quantity.

\subsection{Machine Learning for Pulse Shape Analysis}

Machine learning has been explored for HPGe pulse-shape analysis using both feature-based and end-to-end approaches. Early studies used classical classifiers such as support vector machines and decision trees with hand-crafted waveform features~\cite{BDT_Majorana_Analysis}. More recently, deep learning methods have been applied directly to waveform samples. Convolutional neural networks~\cite{LeCun2015} can learn local temporal patterns but struggle with long-range dependencies due to limited receptive fields, while recurrent neural networks~\cite{Hochreiter1997LSTM} propagate information sequentially but can be inefficient for long sequences.

Transformer architectures~\cite{NIPS2017_3f5ee243} provide an alternative sequence-modelling framework based on self-attention, which relates information across time indices in parallel and can capture both short- and long-range dependencies within a unified mechanism. Attention-based models have shown strong performance across a range of time-series domains~\cite{wen2023transformerstimeseriessurvey}, including other waveform modalities such as medical signals~\cite{ISLAM2024122666}. These properties motivate transformer-based approaches for HPGe pulse-shape analysis, where discriminative cues can involve correlations between distinct regions of the waveform (e.g.,\ rise structure, peak morphology, and late-time behaviour).

A complementary development in machine learning is self-supervised pre-training, where models learn from unlabelled data before fine-tuning on specific tasks~\cite{devlin2019bertpretrainingdeepbidirectional}. Masked autoencoding~\cite{he2021maskedautoencodersscalablevision} exemplifies this approach: portions of the input are randomly hidden, and the model learns to reconstruct the missing content. This forces the encoder to capture general structure in the data distribution. When applied to detector calibration data, masked autoencoding could learn fundamental waveform characteristics, such as charge-collection timescales, noise properties and pulse-shape variations, before specialising to discrimination tasks. This strategy addresses a practical challenge in experimental physics: unlabelled calibration data are abundant, but obtaining labelled training sets for specific analyses requires expert annotation and may be limited by systematic constraints.

\section{Method}

\subsection{Deep learning model}
\label{sec:transformer}

We develop a transformer-based architecture to address the pulse-shape discrimination and energy estimation tasks on digitised waveforms from the \textsc{Majorana Demonstrator} Data Release~\cite{MD_Data_Release}. The transformer encoder~\cite{NIPS2017_3f5ee243}, originally designed for natural language processing, has demonstrated strong performance across diverse sequence modelling domains. Its self-attention mechanism naturally accommodates the temporal structure of detector waveforms without requiring fixed receptive fields, making it well-suited for capturing both local pulse characteristics and long-range dependencies in the signal evolution.

We compare two training strategies: (1) end-to-end supervised learning from labelled waveforms, and (2) a two-stage approach combining self-supervised pre-training on unlabelled data via masked autoencoding, followed by supervised fine-tuning. Both strategies employ the same core transformer architecture, differing primarily in weight initialisation and training objectives. The complete implementation of our model, training procedures, and data processing pipeline is publicly available at \url{https://github.com/mababicz/waveform_mae_hpge}.

\subsubsection{Input Representation and Preprocessing}

Raw waveforms consist of 3{,}800 samples and correspond to digitised preamplifier output charge pulses, rather than deconvolved physical current waveforms. Due to the MJD waveform structure, the rising edge is fully sampled, while the baseline region and the decaying tail are pre-summed every 4 samples and represented in the released traces via linear interpolation in multisampling runs. The dataset consists of 1{,}040{,}000 training waveforms and 390{,}000 test waveforms. We first apply baseline subtraction using the mean of the pre-trigger region to remove detector-specific DC offsets. Beyond the raw baseline-corrected waveform, we compute its first-order gradient via finite differences to provide a simple current-proxy (time derivative of the digitised charge pulse). This dual representation (amplitude and temporal derivative) provides complementary perspectives on pulse shape evolution.

During training, we augment the data by applying random temporal shifts in the range $[-5, +5]$ time steps, using edge padding to maintain the original sequence length. This augmentation improves model robustness to timing variations and slight misalignments in the trigger position. All waveforms and gradients are then standardised using dataset-wide mean and standard deviation statistics computed on the training partition. Similarly, energy labels are normalised to have zero mean and unit variance to stabilise regression training.

We partition each 3,800-step sequence into non-overlapping windows of $W=10$ consecutive time steps, yielding $L=380$ temporal segments per event. Each window spans 100~ns of detector activity. The windowed waveform and gradient sequences are independently projected through learned linear transformations into a shared $d$-dimensional embedding space with $d=64$:
\begin{equation}
    \mathbf{e}_i = \mathbf{W}_{\text{wf}} \mathbf{x}_i^{\text{wf}} + \mathbf{W}_{\text{grad}} \mathbf{x}_i^{\text{grad}},
\end{equation}
where $\mathbf{x}_i^{\text{wf}}, \mathbf{x}_i^{\text{grad}} \in \mathbb{R}^{W}$ denote the waveform and gradient values in window $i$, and $\mathbf{W}_{\text{wf}}, \mathbf{W}_{\text{grad}} \in \mathbb{R}^{d \times W}$ are trainable projection matrices. The resulting embedding $\mathbf{e}_i \in \mathbb{R}^{d}$ integrates both signal characteristics at each temporal location. Following the projection, layer normalisation is applied to stabilise the distribution of embeddings, particularly beneficial given the wide range of energy depositions encountered in calibration data.

Because different detectors exhibit distinct pulse-shape characteristics due to variations in geometry, impurity profiles, and operating conditions, we condition the model on the detector identity. Each of the 26 detectors in the dataset is assigned a learned embedding that produces a pair of per-dimension scale and shift parameters $(\boldsymbol{\gamma}, \boldsymbol{\beta}) \in \mathbb{R}^{d}$ via Feature-wise Linear Modulation (FiLM)~\cite{perez2018film}. After layer normalisation, the token embeddings are modulated element-wise as $\mathbf{e}_i \leftarrow (1 + \boldsymbol{\gamma})\odot \mathbf{e}_i + \boldsymbol{\beta}$, allowing the encoder to adapt its input representation to each detector without requiring separate model parameters. The FiLM parameters are initialised to zero so that the conditioning has no effect at the start of training. Additionally, a per-detector learned affine correction (scale and offset) is applied to the energy regression output, accounting for residual detector-to-detector calibration differences.

\subsubsection{Transformer Encoder Architecture}

The transformer encoder processes the sequence of normalised embeddings through $N=6$ stacked encoder layers. Each layer applies multi-head self-attention followed by a position-wise feed-forward network, both wrapped with residual connections and layer normalisation~\cite{Ba2016LayerNorm} in a pre-norm configuration. Since the model operates on sequential data with inherent temporal ordering, we add sinusoidal positional encodings~\cite{NIPS2017_3f5ee243} to the input embeddings before the first layer, ensuring that positional information is preserved throughout the network.

The multi-head self-attention mechanism with $H=8$ heads enables the model to capture dependencies across different temporal positions and representation subspaces. Each attention head operates on $d/H=8$ dimensions and computes scaled dot-product attention:
\begin{equation}
    \text{Attention}(\mathbf{Q}, \mathbf{K}, \mathbf{V}) = \text{softmax}\left(\frac{\mathbf{Q}\mathbf{K}^\top}{\sqrt{d_k}}\right) \mathbf{V},
\end{equation}
where $\mathbf{Q}$, $\mathbf{K}$, and $\mathbf{V}$ denote the query, key, and value matrices derived from learned linear projections of the layer input. The attention weights dynamically determine which temporal regions are most relevant for characterising each window's context.

The position-wise feed-forward network consists of two linear transformations with a GELU activation function, expanding the representation to an intermediate dimension of $4d=256$ before projecting back to $d$ dimensions. For supervised training, we incorporate stochastic depth (DropPath)~\cite{Huang2016StochasticDepth} with a linearly increasing drop probability reaching $p=0.1$ at the final layer. This regularisation technique randomly drops entire residual paths during training, encouraging the network to develop more robust feature hierarchies without adding computational cost during inference.

Formally, each encoder layer applies:
\begin{align}
    \mathbf{z}_i' &= \mathbf{z}_i + \text{DropPath}\left(\text{MHA}(\text{LN}(\mathbf{z}_i))\right), \\
    \mathbf{z}_{i+1} &= \mathbf{z}_i' + \text{DropPath}\left(\text{FFN}(\text{LN}(\mathbf{z}_i'))\right),
\end{align}
where $\mathbf{z}_i$ represents the hidden state at layer $i$, LN denotes layer normalisation, MHA is multi-head attention, and FFN is the feed-forward network.

\subsubsection{Output Pooling and Prediction Heads}

After the final encoder layer, we obtain a sequence of contextualised representations $\{\mathbf{h}_1, \ldots, \mathbf{h}_L\}$ with $\mathbf{h}_i \in \mathbb{R}^{d}$. To generate event-level predictions from this sequence, we apply global average pooling:
\begin{equation}
    \mathbf{h}_{\text{pool}} = \frac{1}{L} \sum_{i=1}^{L} \mathbf{h}_i.
\end{equation}
This aggregation strategy integrates information across the entire waveform duration into a fixed-size representation. The pooled features pass through an additional layer normalisation before being fed to task-specific prediction heads.

For pulse-shape discrimination, we implement four independent binary classification heads corresponding to the dataset-provided accept/reject labels for DCR, high-side AvsE, low-side AvsE, and LQ (defined in Section~\ref{sec:TraditionalPSD}). Each head is a single linear layer mapping the $d$-dimensional pooled representation to a scalar logit. For energy estimation, a separate regression head predicts the normalised calibrated energy via linear projection. During training, binary cross-entropy loss with label smoothing ($\epsilon=0.05$) is applied to the classification tasks, while Huber~\cite{10.1214/aoms/1177703732} is used for energy regression. The total loss combines these objectives with weights of 1.0 for PSD and 0.5 for energy.

\subsubsection{Masked Autoencoder Pre-training}

To exploit the large volume of unlabelled calibration waveforms, we employ MAE pre-training~\cite{he2021maskedautoencodersscalablevision}. The MAE approach learns representations by reconstructing randomly masked portions of the input, forcing the encoder to capture informative patterns in the visible context.

During pre-training, we randomly mask $\rho=50\%$ of temporal windows and process only the visible windows through the encoder. The encoder architecture matches the supervised model: $d=64$ dimensions, $N=6$ layers, and $H=8$ attention heads. Crucially, stochastic depth is not applied during pre-training, simplifying the self-supervised objective and focusing the model on learning robust waveform features rather than task-specific regularisation. The same temporal shift augmentation and normalisation procedures used in supervised training are applied here as well.

The encoded representations of visible windows are passed to a lightweight decoder that reconstructs the masked content. The decoder inserts learned mask tokens at masked positions and processes the full sequence through $N_{\text{dec}}=2$ transformer layers with dimension $d_{\text{dec}}=64$. Separate linear projection heads output reconstructed values for both the waveform and gradient windows. The reconstruction loss is computed exclusively over masked positions:
\begin{equation}
    \mathcal{L}_{\text{MAE}} = \frac{1}{|\mathcal{M}|} \sum_{i \in \mathcal{M}} \left( \|\hat{\mathbf{x}}_i^{\text{wf}} - \mathbf{x}_i^{\text{wf}}\|_2^2 + \|\hat{\mathbf{x}}_i^{\text{grad}} - \mathbf{x}_i^{\text{grad}}\|_2^2 \right),
\end{equation}
where $\mathcal{M}$ denotes the set of masked window indices and $\hat{\mathbf{x}}_i$ represents the decoder's reconstruction.

After pre-training for 200 epochs, the decoder is discarded and the encoder weights initialise the supervised model. Fine-tuning proceeds end-to-end on labelled data with the combined PSD and energy objectives, now including stochastic depth regularisation. This two-stage training paradigm allows the model to first develop general-purpose waveform representations from abundant unlabelled data before specialising to the downstream discrimination and regression tasks.

\subsubsection{Training}

Table~\ref{tab:hyperparameters} summarises the complete set of hyperparameters used across all training configurations. Both pre-training and supervised learning share the same core architecture (6 transformer layers with 64-dimensional embeddings and 8 attention heads) but differ in their optimisation strategies and regularisation schemes to suit their respective objectives.

\begin{table}[htb]
\centering
\caption{Hyperparameters for MAE pre-training and supervised training (both from scratch and fine-tuning). The base learning rate is scaled linearly according to the effective batch size~\cite{goyal2018accuratelargeminibatchsgd}: $\text{LR}_{\text{eff}} = \text{LR}_{\text{base}} \times \text{batch size} / 256$.}
\label{tab:hyperparameters}
\begin{tabular}{lcc}
\hline
\textbf{Hyperparameter} & \textbf{Pre-training} & \textbf{Supervised} \\
\hline
\multicolumn{3}{c}{\textit{Architecture}} \\
\hline
Embedding dimension ($d$) & 64 & 64 \\
Number of layers ($N$) & 6 & 6 \\
Attention heads ($H$) & 8 & 8 \\
Feed-forward dimension & 256 & 256 \\
Window size ($W$) & 10 & 10 \\
Sequence length ($L$) & 380 & 380 \\
Decoder layers ($N_{\text{dec}}$) & 2 & --- \\
Decoder dimension ($d_{\text{dec}}$) & 64 & --- \\
Mask ratio ($\rho$) & 0.5 & --- \\
\hline
\multicolumn{3}{c}{\textit{Regularisation}} \\
\hline
Stochastic depth (DropPath) & 0.0 & 0.1 \\
Dropout & 0.0 & 0.0 \\
Label smoothing ($\epsilon$) & --- & 0.05 \\
\hline
\multicolumn{3}{c}{\textit{Optimisation}} \\
\hline
Optimiser & AdamW & AdamW \\
Base learning rate & $1.5 \times 10^{-4}$ & $1.0 \times 10^{-3}$ \\
Weight decay & 0.05 & 0.05 \\
$\beta_1$ & 0.9 & 0.9 \\
$\beta_2$ & 0.95 & 0.999 \\
Warmup epochs & 20 & 5 \\
Total epochs & 200 & 20 (fine-tune), 40 (scratch) \\
Batch size & 512 & 256 \\
Gradient clipping & --- & 0.5 \\
Layer-wise LR decay & --- & 0.75 (fine-tune only) \\
\hline
\multicolumn{3}{c}{\textit{Training}} \\
\hline
Train/val split & 0.9 & 0.9 \\
Early stopping patience & 200 & 10 \\
Precision & FP16 & FP16 \\
Temporal shift augmentation & $\pm 5$ steps & $\pm 5$ steps \\
\hline
\multicolumn{3}{c}{\textit{Loss weights}} \\
\hline
PSD classification & --- & 1.0 \\
Energy regression & --- & 0.5 \\
\hline
\end{tabular}
\end{table}

All models are trained using the AdamW optimiser~\cite{Loshchilov2019AdamW}, which combines adaptive learning rates with decoupled weight decay regularisation. The weight decay coefficient of 0.05 provides consistent regularisation across all training scenarios. The momentum parameters differ between pre-training ($\beta_1=0.9$, $\beta_2=0.95$) and supervised training ($\beta_1=0.9$, $\beta_2=0.999$) to balance gradient responsiveness with training stability appropriate for each task.

The learning rate schedule incorporates a linear warmup phase followed by cosine annealing. The base learning rates listed in Table~\ref{tab:hyperparameters} are scaled linearly based on the effective batch size to maintain consistent optimisation dynamics across different hardware configurations: $\text{LR}_{\text{eff}} = \text{LR}_{\text{base}} \times B_{\text{eff}} / 256$, where $B_{\text{eff}}$ is the product of the per-device batch size and the number of GPUs. During MAE pre-training, the learning rate increases from zero to its scaled peak value over 20 epochs, then decays following a cosine schedule until epoch 200. For supervised training, we use different training budgets for the two initialisation strategies: fine-tuning from the pre-trained encoder is run for 20 epochs, with a 5-epoch warmup followed by cosine decay over the remaining 15 epochs, whereas training from scratch is run for 40 epochs to allow additional time for the model to adapt from random initialisation. When fine-tuning from pre-trained weights, we apply layer-wise learning rate decay with a factor of 0.75, assigning progressively smaller learning rates to earlier encoder layers. This strategy preserves the general features captured during pre-training while allowing the task-specific heads and later layers to adapt more readily to the supervised objectives.

We partition the available waveforms into training and validation sets using a 90/10 split. Early stopping monitors validation loss with a patience of 10 epochs for supervised training, preventing overfitting to the training distribution. During pre-training, patience is set to 200 epochs (effectively disabled) since the unsupervised objective benefits from extended training. All training employs mixed-precision (FP16) computation to accelerate convergence and reduce memory consumption. Gradient clipping with a maximum norm of 0.5 is applied during supervised training to prevent instability from occasional large gradients, particularly important when fine-tuning from pre-trained weights.

\subsection{Gradient-Boosted Decision Tree Baseline}
\label{sec:gbdt}

To establish a baseline comparison with classical machine learning methods, we implement a gradient-boosted decision tree (GBDT) classifier for the same PSD tasks~\cite{Friedman2001,Chen2016XGBoost}. GBDTs build strong predictive models by sequentially combining multiple weak learners; in this case, shallow decision trees. Each tree in the ensemble is trained to correct the residual errors of the previous trees, with regularisation terms controlling model complexity to prevent overfitting.

Unlike the transformer architecture, which operates directly on the raw time-series waveforms through learned representations, the GBDT approach requires explicit feature engineering. We extract 12 hand-crafted geometric features (inspired by~\cite{BDT_Majorana_Analysis}) from each waveform that capture key pulse shape characteristics relevant to discrimination tasks:

\begin{itemize}
    \item \textbf{Charge features:} maximum normalised charge ($Q_{\text{max}}$), time index of maximum charge ($Q_{t_{\text{max}}}$), and integrated charge from trigger to end ($A_0$).
    \item \textbf{Current features:} maximum normalised current ($A_{\text{max}}$), time index of maximum current ($A_{t_{\text{max}}}$), and the pulse-to-tail ratio (PTR), defined as the ratio of integrated current before and after the maximum.
    \item \textbf{Derivative features:} maximum normalised jerk ($J_{\text{max}}$), time index of maximum jerk ($J_{t_{\text{max}}}$), and normalised linear approximation of the decay tail ($\delta$).
    \item \textbf{Timing features:} time indices when the waveform reaches 50\% ($t_{50}$) and 99\% ($t_{99}$) of maximum charge, together with the 50--99\% rise time ($t_{\mathrm{rise}}$), which summarises the leading-edge development of the pulse.
\end{itemize}

All amplitude-based features are normalised by the integrated charge $A_0$ to provide scale-invariant representations across different energy depositions. In addition to these 12 geometric features, the detector index is included as an input feature, providing the GBDT with the same auxiliary information available to the transformer through its FiLM conditioning layers. These features encode domain knowledge about pulse shape evolution but are fixed representations that cannot adapt during training, in contrast to the learned embeddings in the transformer architecture.

The GBDT models are trained using standard gradient boosting with binomial log-likelihood loss for binary classification tasks. Separate models are trained for each of the four PSD cuts (DCR, high AvsE, low AvsE, LQ), covering the same set of PSD targets as the transformer, but using separate per-task models without shared representations. Each ensemble consists of 500 trees of depth~5, trained with a learning rate of~0.1, stochastic subsampling of 80\% of the data per iteration, and $\sqrt{p}$ random feature selection per split. Additional regularisation is imposed through minimum sample thresholds of 100 for internal splits and 50 for leaf nodes. While GBDTs remain competitive on tabular data~\cite{GBDT_tabular_data}, they lack the ability to learn hierarchical representations from raw sequential data. This classical approach provides a strong baseline that leverages decades of experience in feature design for detector physics, allowing us to assess the added value of end-to-end learned representations in the transformer models.

\section{Results and Discussion}

We evaluate the transformer-based model (Sec.~\ref{sec:transformer}) on the MJD waveform data using two training strategies: supervised training from scratch and fine-tuning after masked autoencoder pre-training. Our central aim is not only to test whether transformer models can outperform a feature-based baseline, but also to determine whether self-supervised pre-training yields more label-efficient learning from detector waveforms. We therefore compare performance across PSD classification and energy reconstruction tasks, and we explicitly study how model quality scales with the amount of labelled data and the available supervised training budget. For PSD classification, we benchmark against the GBDT baseline introduced in Sec.~\ref{sec:gbdt} to assess the value of direct waveform modelling and learned representations relative to traditional hand-crafted features.

\subsection{PSD label classification}

\begin{figure}[htb]
    \centering
    \includegraphics[width=0.95\linewidth]{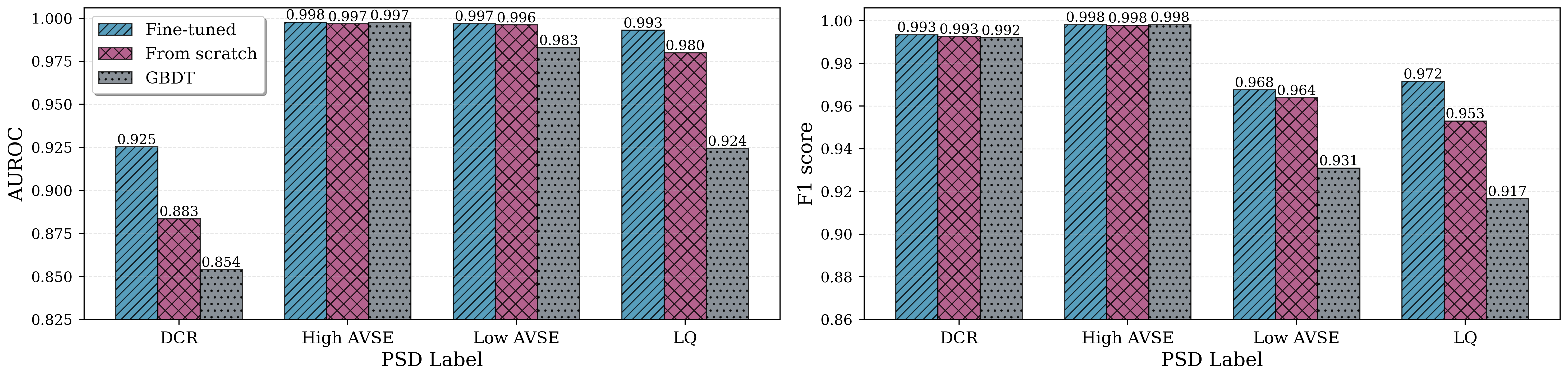}
    \caption{Classification performance comparison for four PSD labels: DCR, high AvsE, low AvsE, and LQ. Both AUROC (left) and F1 score (right) metrics are shown for models trained from scratch (crosshatch pattern) and fine-tuned after MAE pre-training (diagonal hatch pattern). The fine-tuned model achieves consistently higher performance across all labels, with the most pronounced improvement observed for the LQ cut.}
    \label{fig:psd_classification}
\end{figure}

Figure~\ref{fig:psd_classification} presents the classification performance for the four PSD cuts applied in the MJD dataset, comparing both transformer-based approaches against a traditional GBDT baseline. We report two complementary metrics. The area under the receiver operating characteristic curve (AUROC) measures how well a model separates PSD-pass and PSD-fail events independent of any specific operating point or decision threshold; equivalently, it quantifies the probability that a randomly chosen PSD-pass event is assigned a higher score than a randomly chosen PSD-fail event. We additionally report the F1 score at a fixed decision threshold of 0.5, which captures performance at a concrete working point by balancing precision (purity of the selected PSD-pass sample) and recall (PSD-pass efficiency). Using both metrics allows us to distinguish improvements in overall separability (AUROC) from changes that matter for a particular selection threshold (F1), which is directly relevant for downstream event cuts.

Comparing the transformer models to the GBDT baseline in Fig.~\ref{fig:psd_classification} reveals a clear performance gap across all PSD labels, most pronounced for the more challenging cuts. For DCR, transformers achieve substantially higher separability, with AUROC reaching 0.925 (fine-tuned) and 0.883 (scratch) versus 0.854 for GBDT; at the fixed threshold, all three methods have very similar F1 (0.993/0.993 versus 0.992), indicating that the main advantage for DCR appears in ranking performance rather than at this particular operating point. For LQ, the transformer advantage is large in both metrics: AUROC improves to 0.993/0.980 (transformers) compared to 0.924 (GBDT), and F1 increases to 0.972/0.953 compared to 0.917. The AvsE cuts are closer to saturation for all models, but transformers still match or exceed GBDT: for High AvsE, AUROC is 0.998/0.997 versus 0.997 with essentially identical F1 (0.998), while for Low AvsE the AUROC is 0.997/0.996 versus 0.983 and F1 is 0.968/0.964 versus 0.931. Overall, the transformer models provide consistently better class separation (AUROC), with the largest practical gains (F1 at threshold 0.5) appearing for LQ and Low AvsE.

When isolating the effect of the transformer training strategy, both variants demonstrate consistently strong performance, with generally marginal differences across most PSD labels. For the High and Low AvsE cuts, performance is near-optimal for both approaches; AUROC scores differ negligibly (High AvsE: 0.998 vs.\ 0.997; Low AvsE: 0.997 vs.\ 0.996), as do the corresponding F1 scores (High AvsE: 0.998 vs.\ 0.998; Low AvsE: 0.968 vs.\ 0.964). 
Nevertheless, fine-tuning confers clear benefits in the more challenging DCR and LQ settings. For DCR, fine-tuning improves AUROC from 0.883 (from scratch) to 0.925 (+4.2\%), while F1 remains stable at 0.993 for both models, suggesting improved ranking and calibration without materially affecting the operating-point performance. For the LQ cut, fine-tuning increases AUROC from 0.980 to 0.993 (+1.3\%) and improves F1 from 0.953 to 0.972 (+1.9\%). Overall, fine-tuning provides modest-to-moderate gains, with the largest improvements appearing where training from scratch leaves more headroom.

\begin{figure}[htb]
  \centering
  \includegraphics[width=0.9\linewidth]{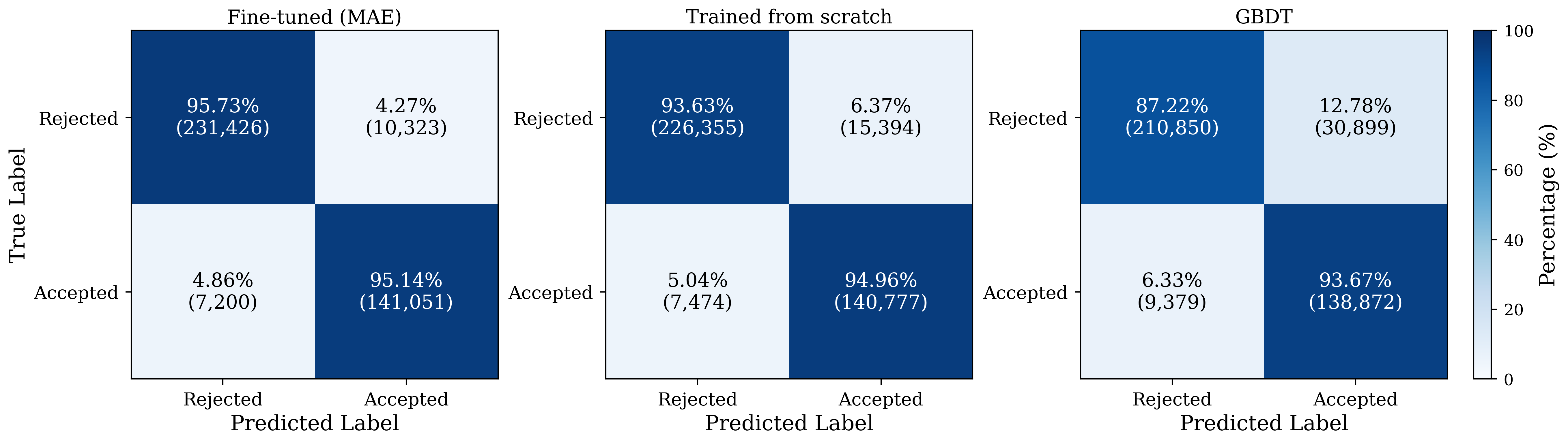}
  \par\bigskip
  \includegraphics[width=0.9\linewidth]{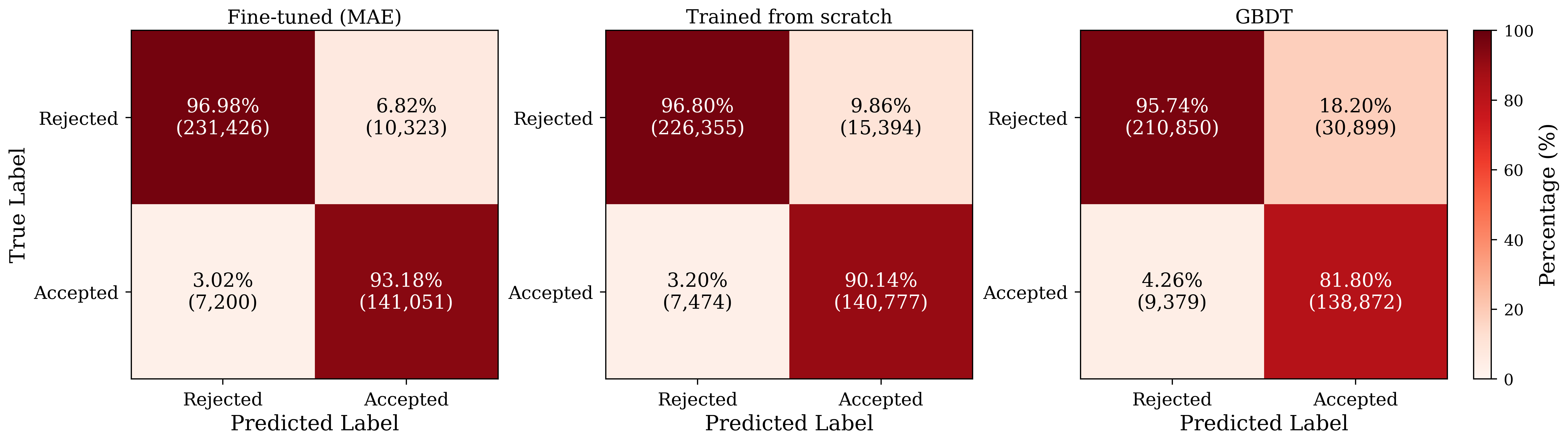}  
  \caption{
    Binary PSD-pass (accepted) versus PSD-fail (rejected) classification performance shown as confusion matrices
    (values are percentages; the colour scale indicates percentage).
    An event is labelled PSD-pass only if it passes all four collaboration-provided PSD accept/reject labels simultaneously
    (low AvsE, high AvsE, DCR, and LQ).
    Each row of panels compares three models: fine-tuned transformer, transformer trained from scratch, and GBDT baseline.
    The top matrices are normalised by true label (row-normalised), highlighting per-class recall, while the bottom matrices are normalised by predicted label (column-normalised), highlighting precision.
  }
  \label{fig:psd_binary_confusion_matrices}
\end{figure}

\begin{table}[htb]
\centering
\caption{Binary PSD-pass versus PSD-fail classification performance. An event is labelled PSD-pass only if it passes all four collaboration-provided PSD accept/reject labels simultaneously (low AvsE, high AvsE, DCR, and LQ).
}
\vspace{0.2cm}
\label{tab:binary_classification}
\begin{tabular}{lccc}
\hline
\textbf{Metric} & \textbf{Fine-tuned (MAE)} & \textbf{Trained from scratch} & \textbf{GBDT} \\
\hline
Accuracy           & \textbf{0.9551} & 0.9414 & 0.8967 \\
Balanced accuracy  & \textbf{0.9544} & 0.9430 & 0.9045 \\
Precision          & \textbf{0.9318} & 0.9014 & 0.8180 \\
Recall             & \textbf{0.9514} & 0.9496 & 0.9367 \\
F1 Score           & \textbf{0.9415} & 0.9249 & 0.8733 \\
AUROC              & \textbf{0.9918} & 0.9866 & 0.9598 \\
\hline
\end{tabular}
\end{table}

These results show that direct waveform modelling with a detector-conditioned transformer provides a clear advantage over a feature-based GBDT baseline for individual PSD cuts. However, the combined PSD-pass versus PSD-fail task reveals additional patterns, as shown in Fig.~\ref{fig:psd_binary_confusion_matrices} and Tab.~\ref{tab:binary_classification}. In particular, the confusion matrices in Fig.~\ref{fig:psd_binary_confusion_matrices} show that both transformer models achieve high and well-balanced per-class recall under row normalisation, but fine-tuning reduces residual failure modes relative to training from scratch: the PSD-fail (rejected) recall increases from 93.6\% to 95.7\%, while the PSD-pass (accepted) recall changes only marginally (95.0\% to 95.1\%). Under column normalisation (precision), the improvement from fine-tuning is more pronounced for the PSD-pass class, increasing accepted precision from 90.1\% to 93.2\% (i.e., reducing the fraction of true rejected events among predicted accepted from 9.9\% to 6.8\%). This trend is reflected in Tab.~\ref{tab:binary_classification}, where fine-tuning leads to higher overall accuracy (0.9551 vs.\ 0.9414), F1 score (0.9415 vs.\ 0.9249), and AUROC (0.9918 vs.\ 0.9866). Both transformer variants substantially outperform the GBDT baseline across these metrics (e.g., accuracy 0.8967, F1 0.8733, AUROC 0.9598), consistent with the larger off-diagonal confusion-matrix entries for GBDT.

\subsection{Energy reconstruction}

\begin{figure}[htb]
    \centering
    \includegraphics[width=0.6\linewidth]{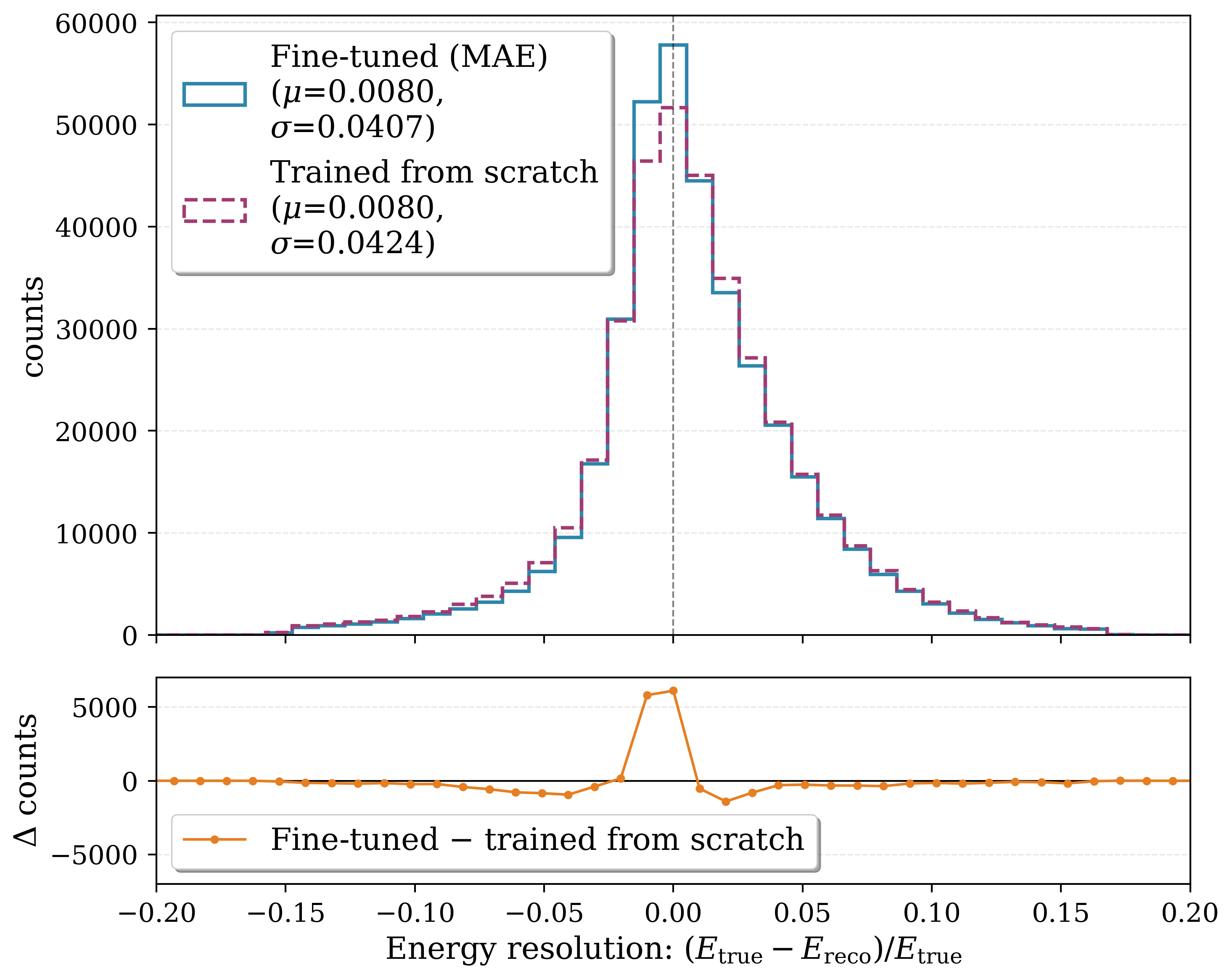}
    \caption{Distribution of the relative residual, defined as $(E_{\mathrm{label}} - E_{\mathrm{pred}})/E_{\mathrm{label}}$, for both training approaches. The top panel shows the histograms for fine-tuned (solid blue) and scratch-trained (dashed magenta) models. Statistical parameters ($\mu$ and $\sigma$) quantify the central tendency and spread of each distribution. The bottom panel displays the difference in counts between the two approaches, highlighting that fine-tuning produces more predictions near zero residual.}
    \label{fig:energy_resolution}
\end{figure}

Energy reconstruction performance is evaluated using the distribution of the relative residual in Fig.~\ref{fig:energy_resolution}. We define the relative residual as $(E_{\mathrm{label}}-E_{\mathrm{pred}})/E_{\mathrm{label}}$, where $E_{\mathrm{label}}$ is the calibrated-energy regression target provided in the dataset. The mean residual is slightly positive for both training strategies ($\mu = 0.0080$ in each case), indicating a small, common tendency to underestimate energies at the $\sim$0.8\% level. The primary difference is instead in the spread: the fine-tuned model exhibits a modestly narrower distribution ($\sigma = 0.0407$) than the scratch-trained model ($\sigma = 0.0424$), indicating slightly tighter agreement with the provided calibrated-energy label. This behaviour is also visible in the lower-panel difference plot (fine-tuned minus scratch), which shows an excess of events near zero residual and corresponding deficits in the surrounding bins, indicating that fine-tuning concentrates more predictions closer to the correct energy.

\begin{figure}[htb]
    \centering
    \includegraphics[width=0.9\linewidth]{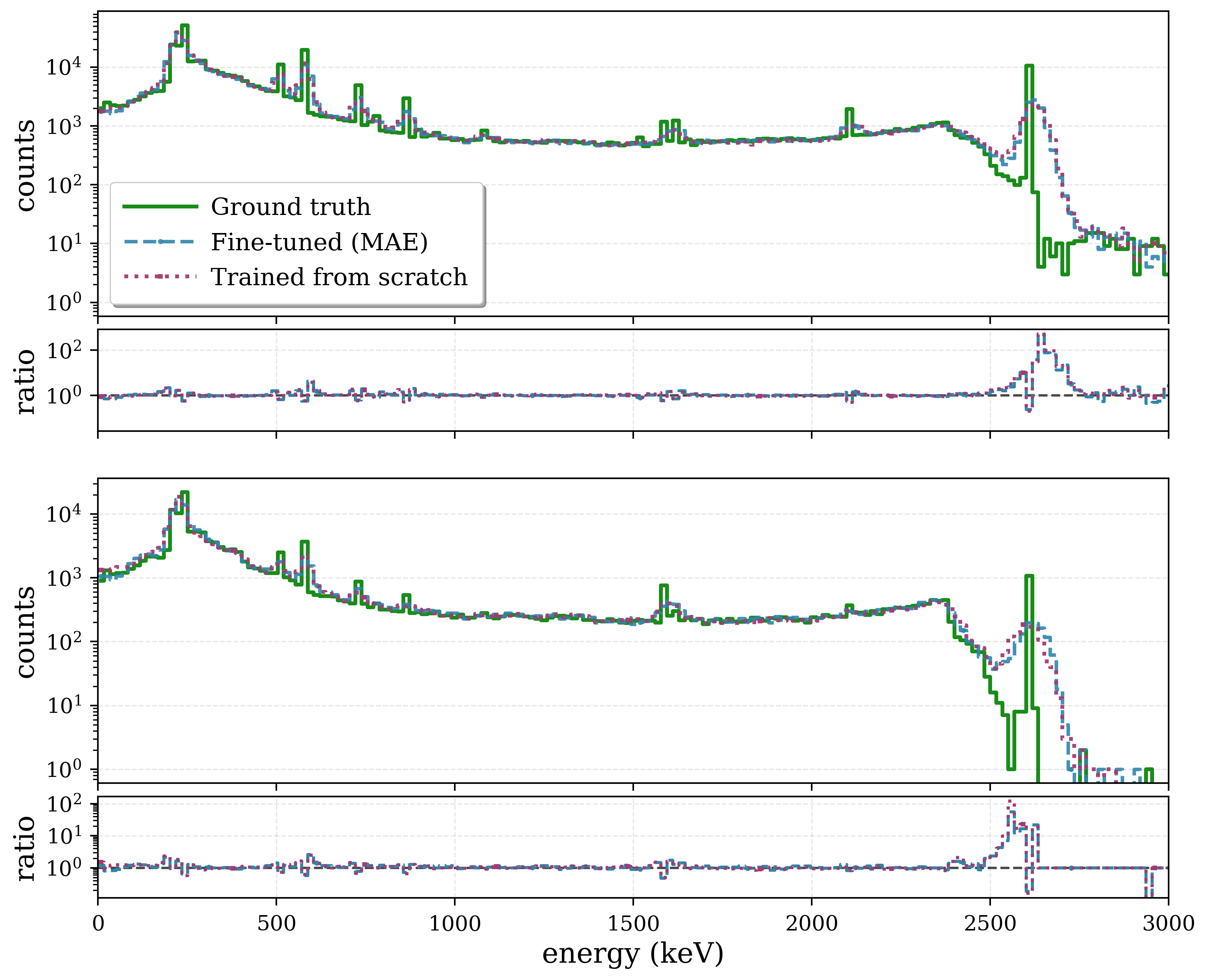}
    \caption{Energy reconstruction performance. Energy spectra on a logarithmic scale. The top panel shows the energy spectra for all events, while the bottom panel shows the PSD-pass subset. For the ground truth curves, PSD-pass events are selected using the true collaboration-provided PSD labels; for the fine-tuned and scratch models, PSD-pass events are selected using the corresponding model predictions. In each case, a ratio panel is shown beneath the spectrum, displaying the model-to-truth ratios.}
    \label{fig:energy_spectrum}
\end{figure}

Figure~\ref{fig:energy_spectrum} provides a complementary, event-level view of the energy regression.
It compares energy spectra on a logarithmic scale.
For the inclusive sample (top plot), the spectra produced using $E_{\mathrm{pred}}$ broadly track the ground-truth shape, including prominent line features, suggesting that the regression does not grossly distort the overall spectral distribution. For the PSD-pass subset (bottom plot), the spectra obtained after applying model-predicted selections follow the same qualitative trends as the spectrum obtained from the collaboration-provided labels, with differences most apparent in regions where the PSD-pass spectrum becomes sparse at the high-energy end. Overall, Figure~\ref{fig:energy_spectrum} indicates that both training approaches yield similar global behaviour, while also highlighting the presence of a secondary response band that warrants explicit treatment (e.g., stratified evaluation) in downstream analyses.

\subsection{Training epochs vs training data}

To evaluate the sample efficiency of the pre-training approach, we vary both the number of supervised training epochs (2, 5, 10, and 20) and the amount of labelled training data (65K, 130K, 260K, 520K, and 1,040K waveforms). For each configuration, we compare models trained from scratch against models initialised with the pre-trained encoder. This systematic comparison isolates the practical value of self-supervised learning for HPGe waveform analysis: whether representations learned from abundant unlabelled calibration waveforms can reduce the labelled-data and optimisation budget required for downstream PSD tasks.

\begin{figure}[htb]
    \centering
    \includegraphics[width=1.0\linewidth]{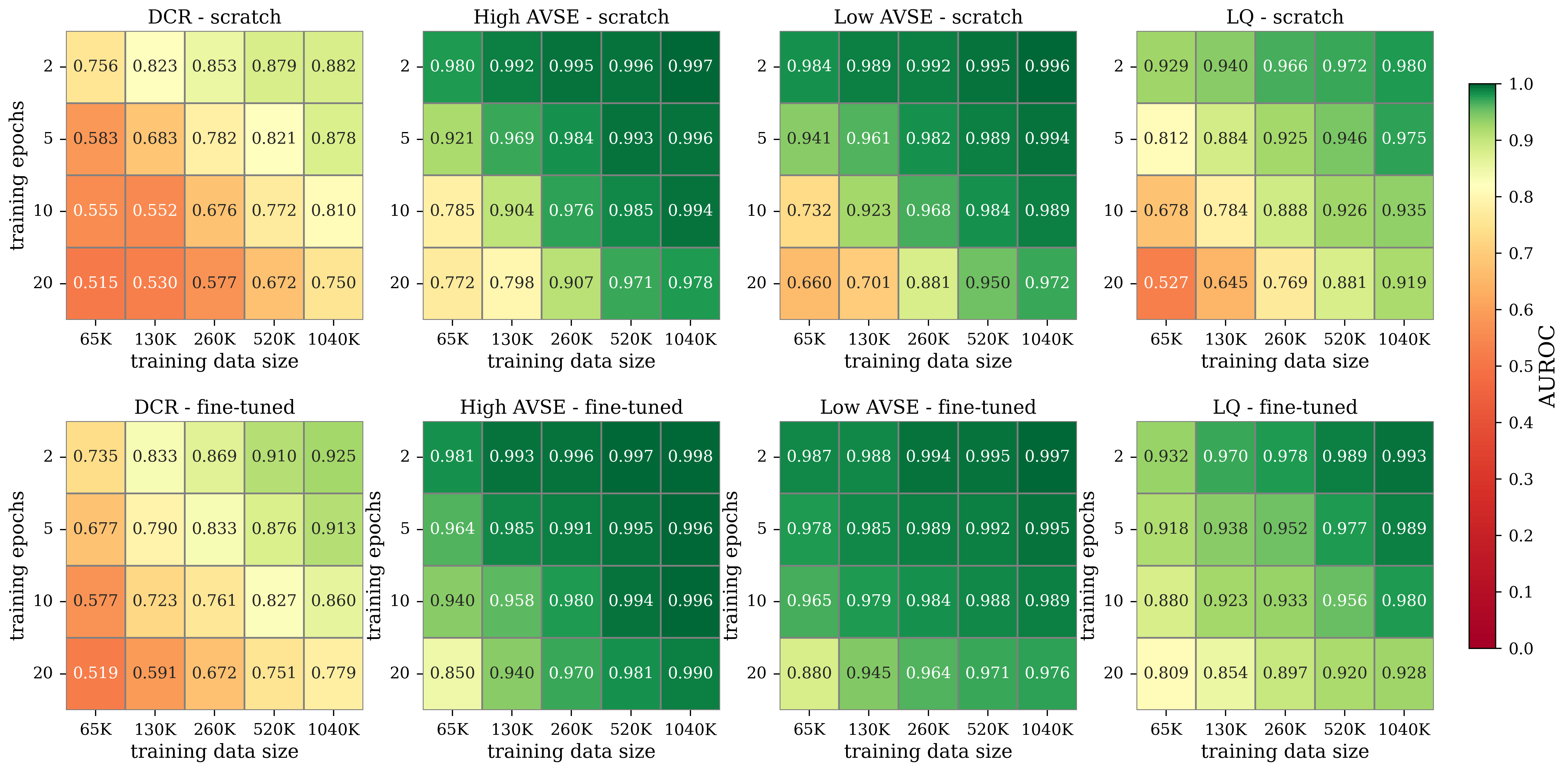}
    \caption{Sample efficiency comparison across training configurations. Classification AUROC for four PSD cuts (DCR, high AvsE, low AvsE, LQ) as a function of training data size (65K--1,040K waveforms) and training duration (2--20 epochs). Top row: models trained from scratch; bottom row: models fine-tuned after MAE pre-training. Warmer colors indicate higher AUROC.}
    \label{fig:heatmap}
\end{figure}

Figure~\ref{fig:heatmap} summarises AUROC as a function of training set size and supervised training budget (epochs) for scratch training (top row) and MAE fine-tuning (bottom row). The clearest effect of pre-training is improved sample efficiency in the low-data/low-epoch regime. For example, with only 65k waveforms and 2 epochs, fine-tuning substantially increases AUROC for several labels: Low AvsE improves from 0.660 (scratch) to 0.880 (fine-tuned) and LQ improves from 0.527 to 0.809. Even at the same small dataset size, the gains persist at 5 epochs (Low AvsE: 0.732 $\rightarrow$ 0.965; High AvsE: 0.785 $\rightarrow$ 0.940; LQ: 0.678 $\rightarrow$ 0.880). In practical terms, fine-tuning can match scratch performance obtained with substantially more labelled data: e.g., for Low AvsE at 2 epochs, 65k fine-tuned (0.880) is comparable to scratch with 260k waveforms (0.881), and for LQ at 5 epochs, 65k fine-tuned (0.880) is comparable to scratch with 260k waveforms (0.888).

As the training budget increases, both strategies approach saturation for the easier AvsE cuts and the pre-training advantage narrows to the $\mathcal{O}(10^{-3})$ level at the largest settings (e.g., High AvsE at 20 epochs: 0.998 vs.\ 0.997; Low AvsE at 20 epochs: 0.997 vs.\ 0.996). In contrast, the more challenging cuts retain a visible gap even at high data/epoch budgets: at 1040k waveforms and 20 epochs, LQ improves from 0.980 (scratch) to 0.993 (fine-tuned), and DCR improves from 0.882 to 0.925. Overall, these heatmaps indicate that MAE pre-training is most valuable for reducing the labelled-data and epoch requirements to reach a target performance, while also yielding the largest absolute improvements on the hardest PSD labels.

These results have practical implications for experimental physics applications. When computational budgets or data availability constrain model development, MAE pre-training offers a viable strategy to extract maximum performance from limited resources. 
The approach is particularly valuable in ongoing analyses and in studies of new detector configurations, where extensive labelled datasets may not yet exist and rapid iteration of PSD strategies is important.
Similar data efficiency gains from masked autoencoding pre-training have been recently demonstrated in other particle physics applications~\cite{Young2025PoLArMAE}, where fine-tuning on 100 labelled Liquid Argon Time Projection Chamber (LArTPC) neutrino events achieved performance comparable to supervised baselines trained on over 100,000 events.

\section{Conclusion}

In this study, we have shown that transformer-based architectures consistently outperform the feature-based GBDT baseline for pulse-shape discrimination in high-purity germanium detectors. Our evaluation on the MJD data release shows that the largest gains appear for the more challenging PSD targets and for the combined PSD-pass label: for the latter, the fine-tuned transformer improves AUROC from 0.9598 to 0.9918 and F1 from 0.8733 to 0.9415 relative to the GBDT baseline. MAE pre-training further improves sample efficiency, reducing the labelled-data requirement by roughly a factor of 2--4 in the low-data regime, and yields a modest narrowing of the energy-residual distribution. Both transformer variants retain a small common underestimation bias on the provided test split.

These gains are encouraging for $0\nu\beta\beta$ analyses, but for MJD they should not yet be interpreted as a quantified improvement in half-life sensitivity. Translating the classification improvements reported here into an actual $0\nu\beta\beta$ sensitivity gain requires an ROI-level study of signal efficiency, background acceptance, and associated systematic uncertainties under the final analysis selection. Performing that translation is the most important next physics step. More broadly, the relevance of this approach extends beyond MJD. Waveform-based PSD is an important component of several low-background HPGe-detector analyses, so methods that improve PSD performance and reduce labelled-data requirements should be transferable not only to the ongoing \textsc{LEGEND}-200 analysis and the future \textsc{LEGEND}-1000 program, but also to other HPGe-based experiments where background rejection is critical. Future work should test robustness across detectors, runs, and calibration conditions; evaluate energy-dependent performance in the $Q_{\beta\beta}$ region of interest; and propagate the resulting signal-efficiency and background-rejection changes into a full $0\nu\beta\beta$ sensitivity study.

\ack{}

\appendix

\section{Masked Autoencoder Reconstruction Analysis}

To validate the effectiveness of the masked autoencoder (MAE) pre-training strategy, we analyse the model's reconstruction capabilities on held-out test data. This analysis provides insights into the quality of learned representations and confirms that the encoder captures meaningful waveform features during the self-supervised pre-training phase.

\subsection{Reconstruction Methodology}

The MAE model is trained to reconstruct randomly masked portions of the input waveform and gradient sequences. During pre-training, 50\% of the 380 temporal windows are randomly masked, and the model must predict the values in these masked regions based solely on the visible context. The reconstruction task forces the encoder to learn robust internal representations that capture both local pulse characteristics and global waveform structure.

To evaluate reconstruction quality, we load the full pre-trained MAE model (encoder and decoder) from the checkpoint saved after 200 epochs and apply it to test events that were not seen during training. For each test event, we randomly mask 50\% of windows following the same procedure used during training, then measure the mean squared error (MSE) between the original and reconstructed values within masked regions only. This metric directly reflects the model's ability to infer missing information from partial observations.

\subsection{Qualitative Reconstruction Examples}

We visualise reconstructions for representative events with different interaction topologies to assess whether reconstruction quality depends on event type. The visualisations display the original waveform and gradient signals, with masked regions (50\% of temporal windows) highlighted in blue shading. Within masked regions, the original values are shown as solid blue lines while reconstructed predictions appear as red scatter points. Each figure includes a zoomed inset (green dashed border) showing the rising edge detail, where the steepest signal gradient occurs and accurate reconstruction is most challenging.

\begin{figure}[htb]
    \centering
    \includegraphics[width=0.95\linewidth]{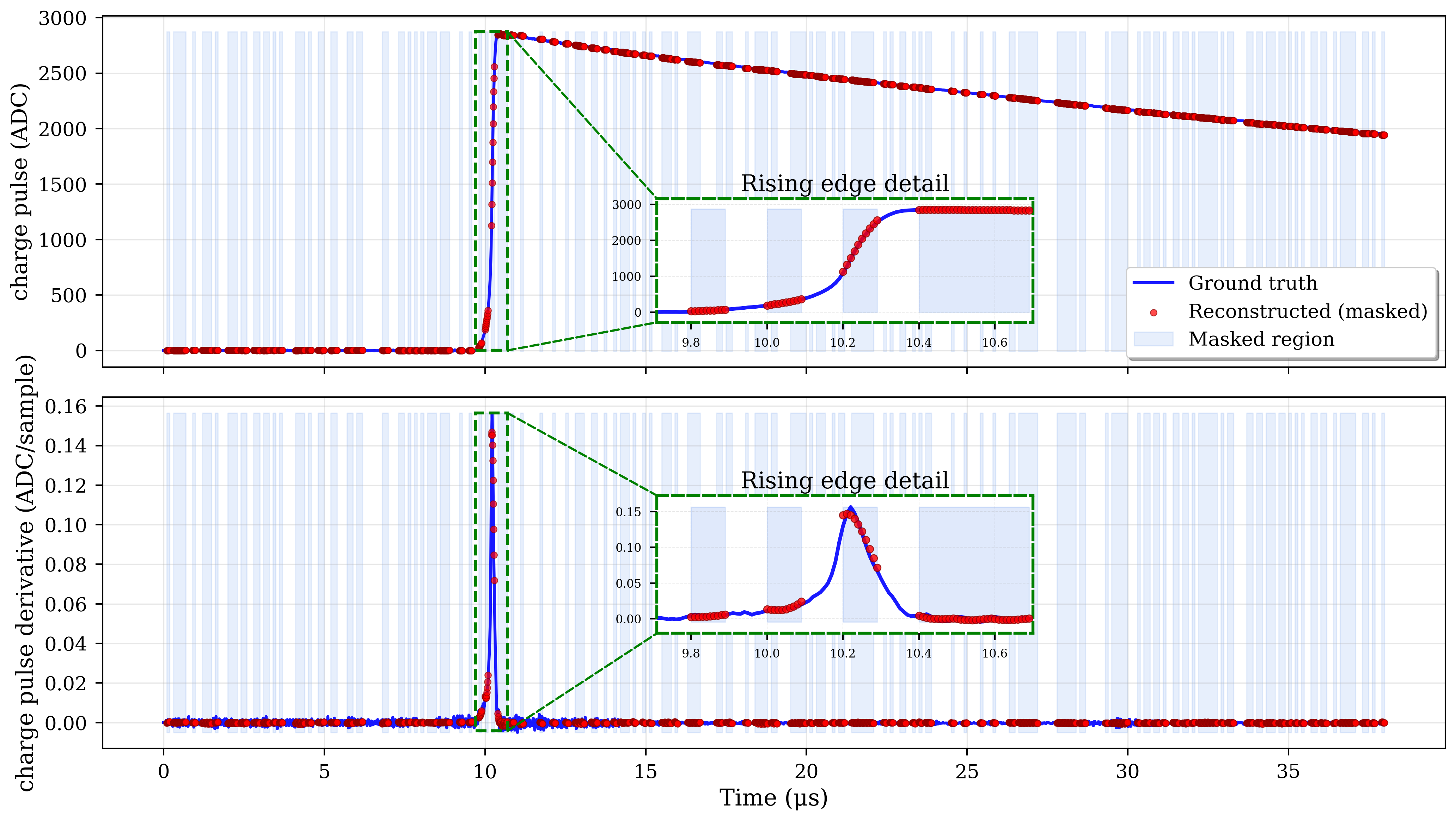}
    \caption{MAE reconstruction for a representative PSD-pass event passing all four collaboration-provided PSD labels. Top: waveform reconstruction showing charge pulse amplitude (ADC units). Bottom: gradient reconstruction showing temporal derivative (ADC/sample). Blue shading indicates masked regions; blue lines show ground truth; red points show model predictions. The zoomed inset (green box) focuses on the rising edge, demonstrating accurate reconstruction even at the steepest gradient. The model captures both global pulse morphology and fine-grained temporal structure.}
    \label{fig:mae_reconstruction_signal}
\end{figure}

Figure~\ref{fig:mae_reconstruction_signal} shows reconstruction for an event that is PSD-pass according to the four collaboration-provided accept/reject labels (DCR, high AvsE, low AvsE, and LQ). In the context of this dataset, PSD-pass indicates agreement with the collaboration PSD selections rather than physics-truth single-site topology. The MAE model accurately reconstructs both the waveform and its gradient in masked regions, with reconstructed points closely following the true signal evolution. The rising edge inset reveals that even the steepest temporal gradient region—where charge collection dynamics are most rapid—is faithfully predicted from the surrounding context, demonstrating high fidelity across the pulse structure.

\begin{figure}[htb]
    \centering
    \includegraphics[width=0.95\linewidth]{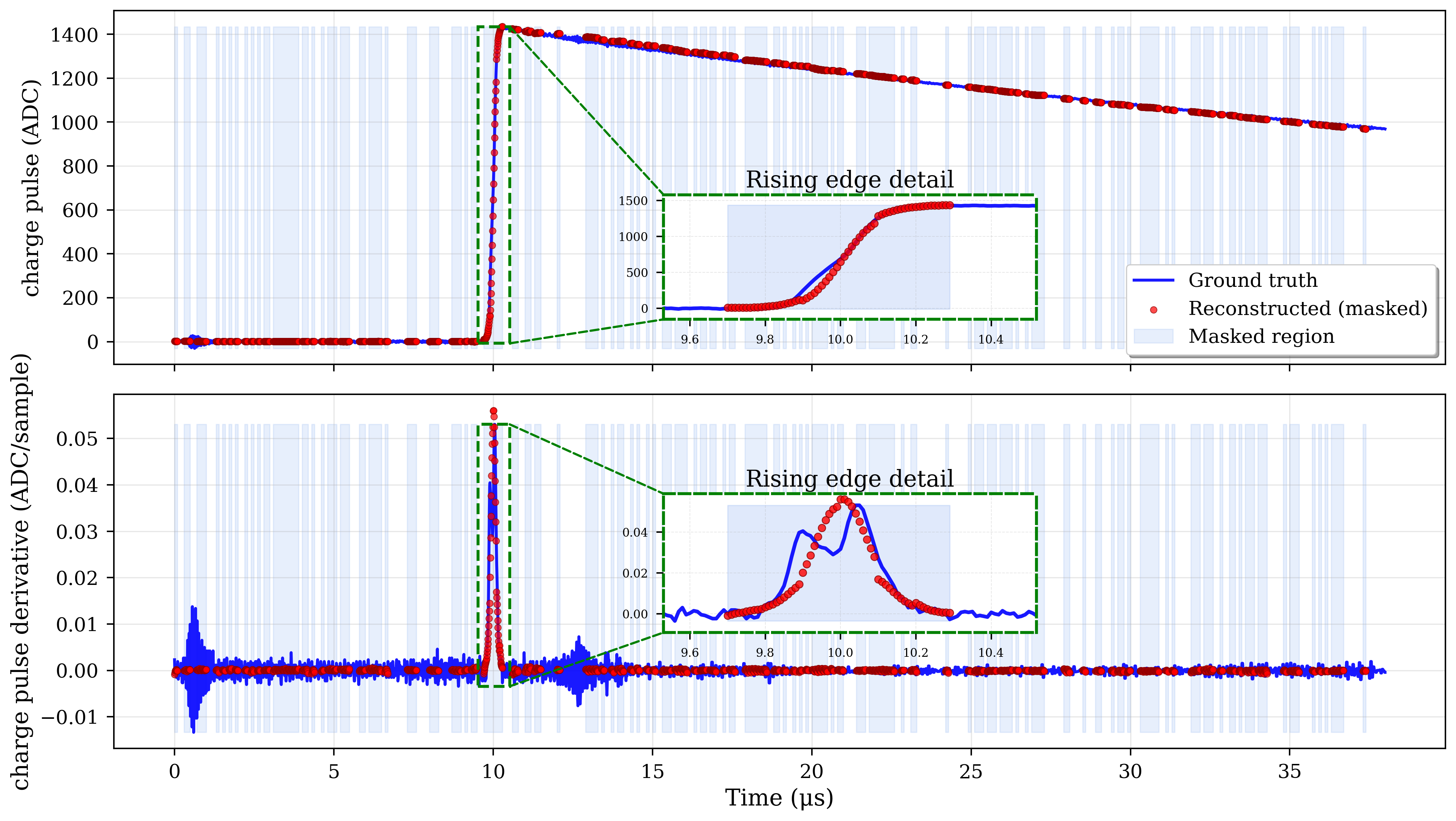}
    \caption{MAE reconstruction for a representative event failing the low AvsE PSD label, consistent with multi-site-like pulse morphology. Top: waveform reconstruction showing charge pulse amplitude (ADC units). Bottom: gradient reconstruction showing temporal derivative (ADC/sample). Blue shading indicates masked regions; blue lines show ground truth; red points show model predictions. The zoomed inset (green box) focuses on the rising edge. The gradient exhibits a double-peak structure (visible in the ground truth) characteristic of spatially separated energy depositions. While the model accurately reconstructs the overall waveform, it approximates the gradient multi-peak structure rather than recovering its precise timing. This limitation is reasonable: inferring such rare, event-specific features from partial context is inherently difficult. The encoder still learns useful general representations of detector response that transfer effectively to downstream tasks.}
    \label{fig:mae_reconstruction_multisite}
\end{figure}

Figure~\ref{fig:mae_reconstruction_multisite} presents reconstruction for an event that fails the low AvsE label while passing DCR, high AvsE, and LQ labels. Multi-site-like events can result from Compton scattering or other processes producing energy depositions at multiple locations, leading to modified pulse shapes compared to single-site-like interactions. The failing low AvsE classification is consistent with an amplitude-to-energy ratio characteristic of such multi-site-like topology. The gradient panel reveals a notable feature of this event: a double-peak structure reflecting the temporal separation between distinct energy depositions. The MAE model does not fully recover the fine details of the double-peak gradient structure. This is expected: predicting such rare, event-specific features from the surrounding context alone is inherently challenging, as the multi-peak signature requires information from the masked regions themselves. Nonetheless, the model captures the dominant waveform characteristics and provides a reasonable prediction given the partial information available.

Across both event types examined, the MAE model demonstrates high-quality reconstruction of general waveform structure. For PSD-pass events, the model closely tracks all aspects of the signal evolution including fine-grained temporal details. For PSD-fail, the model captures the dominant pulse morphology and rising edge dynamics, though it understandably struggles with rare event-specific features like gradient multi-peak structures that require information from the masked regions themselves. Importantly, this behaviour is appropriate for pre-training: the objective is to learn universal representations of detector response—such as charge collection dynamics, noise properties, and typical pulse shapes—rather than memorizing rare event-specific signatures. The learned representations provide strong priors for common waveform characteristics while allowing the downstream supervised model to specialise in distinguishing subtle event-type-specific features during fine-tuning. This explains why transfer learning remains effective: the pre-trained encoder provides a robust initialisation capturing general detector physics, enabling efficient adaptation to PSD and energy tasks even when occasional reconstruction ambiguities arise for unusual event topologies.

\subsection{Quantitative Reconstruction Statistics}

To quantify reconstruction performance systematically, we compute MSE statistics across 1,000 randomly sampled test events. Table~\ref{tab:mae_reconstruction_stats} summarises the reconstruction errors separately for PSD-pass and PSD-fail events.

\begin{table}[htb]
\centering
\caption{Reconstruction error statistics (MSE) for waveform and gradient predictions in masked regions, stratified by PSD-pass/PSD-fail category. Mean and standard deviation are computed across 1,000 test events.}
\label{tab:mae_reconstruction_stats}
\begin{tabular}{lccc}
\hline
\textbf{Channel} & \textbf{All Events} & \textbf{PSD-pass} & \textbf{PSD-fail}
 \\
\hline
Waveform  & $1.2 \times 10^{-4} \pm 0.8 \times 10^{-4}$ & $1.1 \times 10^{-4} \pm 0.7 \times 10^{-4}$ & $1.3 \times 10^{-4} \pm 0.9 \times 10^{-4}$ \\
Gradient  & $1.5 \times 10^{-4} \pm 1.0 \times 10^{-4}$ & $1.4 \times 10^{-4} \pm 0.9 \times 10^{-4}$ & $1.6 \times 10^{-4} \pm 1.1 \times 10^{-4}$ \\
\hline
\end{tabular}
\end{table}

The reconstruction errors are remarkably low across all categories, with mean MSE values on the order of $10^{-4}$ for both waveform and gradient channels. The similarity in reconstruction quality between PSD-pass and PSD-fail categories further confirms that the MAE objective encourages learning of general-purpose representations applicable to diverse waveform morphologies. The slightly higher gradient reconstruction error reflects the increased difficulty of predicting derivative information, which is more sensitive to local variations.

These quantitative results validate the MAE pre-training approach: the model successfully learns to reconstruct masked waveform segments with high fidelity, demonstrating that the encoder captures rich structural information about HPGe detector signals. This learned knowledge provides a strong foundation for transfer to downstream supervised tasks, explaining the performance improvements and sample efficiency gains observed in the main results.

\subsection{Implications for Transfer Learning}

The high-quality reconstructions achieved by the MAE model have important implications for understanding why pre-training improves downstream task performance. The reconstruction task requires the encoder to develop internal representations that capture essential characteristics of waveform evolution, including pulse rise times, decay patterns, and noise properties. These learned features are directly relevant to pulse-shape discrimination, as the temporal structure encoded during pre-training provides a strong prior for distinguishing PSD-pass from PSD-fail events.

Furthermore, the consistency of reconstruction quality across different event types suggests that the pre-trained encoder provides a robust starting point for fine-tuning. Rather than learning waveform fundamentals from scratch during supervised training, the fine-tuned model can leverage pre-existing knowledge about temporal structure and focus its capacity on task-specific discrimination boundaries. This explains both the improved precision observed in PSD tasks and the slightly improved agreement with the provided calibrated-energy target, as the model begins with a more informative feature space.

The successful application of MAE pre-training to detector waveforms demonstrates the potential of self-supervised learning in experimental physics. By exploiting the abundant unlabelled calibration data available in most experiments, masked reconstruction provides a principled approach to learning useful representations before engaging with limited labelled datasets. This methodology is broadly applicable to other detector technologies and measurement modalities where temporal or spatial structure can be predicted from partial observations.

\bibliographystyle{iopart-num}
\bibliography{main}

\end{document}